\documentclass{article}
\fontencoding{T1}
\usepackage{anysize}
\usepackage{cite}
\marginsize{1.8cm}{1.8cm}{1.5cm}{1.5cm}
\usepackage[]{epsfig}
\usepackage{amsfonts}
\usepackage{amsmath}
\usepackage{cite}
\usepackage{pdflscape}

\def\vec#1{\mathbf{#1}}
\def\vr{\vec{r}}
\def\vs{\vec{s}}

\def\vv{\vec{v}}

\def\vxi{\pmb \xi}

\begin{document}
\title{Short-time dynamics in active systems: the Vicsek model}

\author{M.~Leticia {Rubio Puzzo}$^{1,2,3}$, Ernesto S.~Loscar$^{1,2,3}$, Andr\'es De Virgiliis$^{1,2}$, Tom\'as S.~Grigera$^{1,2,3,4}$}

\date{\small$^1$Instituto de F\'isica de L\'iquidos y Sistemas Biol\'ogicos (IFLySiB) --- Universidad Nacional de La Plata and CONICET, Calle 59 n.\ 789, B1900BTE La Plata, Argentina \\$^2$CCT CONICET La Plata, Consejo Nacional de Investigaciones Cient\'ificas y T\'ecnicas, Argentina \\$^3$ Departamento de F\'isica, Facultad de Ciencias Exactas, Universidad Nacional de La Plata, Argentina \\$^4$Istituto dei Sistemi Complessi, Consiglio Nazionale delle Ricerche, via dei Taurini 19, 00185 Rome, Italy}

\maketitle

\begin{abstract}
  We study the short-time dinamycs (STD) of the Vicsek model with vector noise.  The study of STD has proved to be very useful in the determination of the critical point, critical exponents and spinodal points in equilibrium phase transitions.  Here we aim is to test its applicability in active systems.  We find that, despite the essential non-equilibrium characteristics of the VM (absence of detailed balance, activity), the STD presents qualitatively the same phenomenology as in equilibrium systems.  From the STD one can distinguish whether the transition is continuous or discontinuous (which we have checked also computing the Binder cumulant).  When the transition is continuous, one can determine the critical point and the critical exponents.
\end{abstract}

\section{Introduction}

Active matter systems are characterised by their ability to transduce free energy into systematic movement \cite{ramasw}.  Both ``active'' and ``externally driven'' particles require input of energy to generate and sustain movement in a dissipative environment, and in this sense they are similar \cite{menzel}.  However, one can distinguish between energy conversion at the micro-scale (active particles) and energy conversion at the meso- or macro-scale (driven particles) \cite{Bowick}.
The active particle concept is also useful because the energy conversion process
is often sufficiently complicated (e.g.\ in moving animals, from bacteria to mammals) that in modelling behaviour at scales much larger than the particle size it is best to assume the particle moves by itself, without necessarily being perturbed by an external influence.  Thus in simple models of active matter, often devised to model the movement of biological groups, this translates into the statement that particles can set their own speed (possibly with fluctuations) through an unspecified internal mechanism.

Active matter (biological and non-biological) has attracted great attention in recent years, being the focus of many experimental, theoretical and numerical works (see \cite{ramasw, menzel, marchetti_hydrodynamics_2013, reviewVicsek,Bechinger,Chate2020} for reviews).  Its intrinsically non-equilibrium nature creates new challenges for its complete understanding, and requires extending or modifying the concepts and tools developed in statistical mechanics.  As an example of the difficulties posed by these systems we can cite the long discussion around the nature of the order-disorder transition in the Vicsek model \cite{chate,baglietto2,reviewVicsek}. The early literature assumed that the onset of order was very similar to an equilibrium continuous transition, but over the years it became clear that the situation is that of a discontinuous transition with phase separation between a disordered gas and an ordered liquid.  The ordered phase is a dense active liquid with orientational order, and instead of simply two regions (ordered and disordered) in parameter space, there are three, corresponding to homogeneous disordered gas, homogeneous ordered liquid, and in between a coexistence region with dense ordered bands separated by sparse disordered fluid \cite{Chate2020}.

In this paper we consider whether the STD technique can be used in active matter.  STD has proven to be very useful to study equilibrium phase transitions.  It is based on the relaxation to equilibrium of systems with parameters tuned near the ordering transition.  Janssen and collaborators found \cite{janssen1989} that the relaxation of a critical system displays universal features, in particular that this out-of-equilibrium regime can be described with a set of critical exponents comprising the static and dynamic equilibrium exponents, plus an independent one, purely out-of-equilibrium, called the ``initial slip'' exponent (see Sec.~\ref{sec:short-time-dynamics} for a summary).  This work opened the way to the use of STD as a technique to determine the critical control parameter for the transition and the critical exponents in numerical simulations, without the need for thermalization \cite{Zheng, Albano_2011}.  It was later noticed that STD can also be applied for discontinuous transitions \cite{schulke_dynamic_2000}, where it can determine the spinodal points that bracket the discontinuous transition \cite{YassJCP2009, Albano_2011}.

STD is a potentially very useful technique to be added to the toolbox of those trying to understand active matter.  In particular the possibility of determining the location and nature of an ordering transition without thermalizing can be very helpful, allowing for example to reach system sizes that would otherwise be prohibitive.  In this work we study the short-time behaviour of an active system, and we find that, with some care, it is possible to find the ordering transition following essentially the same procedure as in equilibrium systems.

To test STD we use the Vicsek model (VM) \cite{VM}, one of the simplest and better known active models, and described in detail in Sec.~\ref{sec:aclvm}.  The VM is out of equilibrium because of the self-propulsion of its constituent particles, but also because the dynamical rule is specified in a way that does not respect detailed balance (Sec.~\ref{sec:detailed-balance}).  For this reason we study both the static (i.e.\ on-lattice) VM (Sec.~\ref{sec:aclstd-static-aclvm}) and the standard active, off-lattice model (Sec.~\ref{sec:aclstd-active-aclvm}).  We also consider two variations of the dynamical rule (parallel vs.\ sequential update), which turn out to lead to different behaviour.  Our conclusions are summarised in Sec.~\ref{sec:conclusions}.

\section{Model and methods}

\subsection{The Vickse model with vector noise}

\label{sec:aclvm}

The VM consists of $N$ self-propelled particles moving in a $d$-dimensional box of side $L$ with a fixed speed $v_0$ (all simulations in this work are done in $d=3$).  At each time step positions $\vr_i(t)$ and velocities $\vv_i(t)$ are updated according to
\begin{subequations}
  \begin{align}
    \vv_i(t+\Delta t) &= v_0 \mathcal{N}\left[ \sum_{j\in S_i} \vv_j + N_i \hat\eta \vxi \right] \label{vm1} \\
               &= v_0 \mathcal{N} \left[\sum_{j\in S_i} \vs_j + N_i (\hat\eta/v_0) \vxi \right], \\
    \vr_i(t+\Delta t) & = \vr_i(t) + \Delta t \vv_i(t+\Delta t), \label{vm2}
\end{align}
  \label{VMeqs}
\end{subequations}
where $S_i$ is a sphere or radius $r_c$ centred at $\vr_i(t)$ (the interaction sphere), $\mathcal{N}(\vv) = \vv/ \lvert\vv\rvert$, $\vs_i=\vv_i/v_0$, $N_i$ is the number of particles within the interaction sphere of particle $i$ and $\vxi$ is a random vector on the unit sphere.  $\eta=\hat\eta/v_0$ is a real number that controls the intensity of the noise and, together with the density $\rho=N/L^d$, is the parameter that controls the order-disorder transition (note that the static limit $v_0\to0$ is taken with fixed $\eta$).  In passing from \eqref{vm1} to \eqref{vm2} we have simply multiplied the argument of the normalisation operator $\mathcal{N}$ by $v_0^{-1}$, which leaves the result of the operator unchanged. 

At low noise and high density the velocities align and move together in a flock.  The development of this order can be monitored with the polarisation,
\begin{equation}
  \label{eq:OP}
  \Phi = \frac{1}{N} \left\lvert \sum_i \frac{\vv_i}{v_0} \right\rvert.
\end{equation}

The VM as defined by \eqref{VMeqs} was introduced by Chat\'e et al.\ \cite{chate2008}.  It differs from the original proposal by Vicsek et al. \cite{VM} in two aspects: the way noise is introduced (which is a vector noise in the case of \eqref{vm1}, while in the original VM it is an angular noise), and the use of a forward update in \eqref{vm2} as opposed to the backward update used in the original paper.  It is generally accepted that forward or backward update does not introduce important changes in behaviour \cite{ginelli_physics_2015}, although small shifts in $\eta_c$ have been reported \cite{baglietto2}.  In contrast, the choice of noise rule, although not always emphasised, can produce some qualitative changes in behaviour (e.g. \cite{Rubio19}).  For the present work, the most important difference between angular and vector noise is that the discontinuous nature of the transition becomes evident at smaller sizes when using the latter.

Equations~\ref{VMeqs} define a paralell update (PU) dynamics: all new velocities are computed from positions and velocities at time $t$, then all positions are updated simultaneously.  Most of the following results were obtained with PU and $\Delta t=1$.  One may alternatively consider a sequential update (SU), as most often employed in Monte Carlo simulations of spin systems, where one particle is picked at random, its velocity and position is updated, and then another particle is picked at random.  We performed some simulations with SU and $\Delta t=1/N$.

\subsection{Detailed balance in the VM}
\label{sec:detailed-balance}

The VM is an intrinsically out-of-equilibrium model due to the particles' activity, i.e.\ the fact that their speed is not set by the interactions as in physical fluids, but determined externally.  However, if one turns off \eqref{vm2} the model becomes on-lattice, $\vv_i$ reduces to an internal degree of freedom and the dynamic equation \eqref{vm1} can be thought as arising from a ferromagnetic Hamiltonian.  In this sense one can think the $v_0\to0$ limit of the VM is the (classical) Heisenberg model (on some random lattice).  However, equivalence to the Heisenberg model requires a dynamics that asymptotically reaches the corresponding Boltzmann distribution.  In Monte Carlo simulations this is normally ensured by enforcing the detailed balance condition,
\begin{equation}
P_\mu W_{\mu\to \nu} = P_\nu W_{\nu\to \mu}, \label{eq:detbal}
\end{equation}
where $P_\mu$ is the Boltzmann probability of state $\nu$ and  $W_{\mu\to\nu}$ is the transition probability.  The dynamics of the VM does not obey detailed balance with respect to the Boltzmann distribution.  This is easiest to see in the SU case and for small $\eta$: take a configuration that is highly ordered.  When updating the velocity of a selected particle according to~\eqref{vm1}, its velocity is first aligned with the neighbours and then perturbed with a small vector.  Thus the probability of producing a badly misaligned velocity (a ``flipped'' particle) is strictly zero, whereas the transition probability from a configuration with a single misaligned particle to a highly aligned one is finite.  Thus these transition probabilities cannot obey~\eqref{eq:detbal}.

Finally, let us note that there is an additional element in \eqref{vm1} that can lead to non-equilibrium effects even for $v_0\to0$.  We can rewrite \eqref{vm1} as
\begin{equation}
\vv(t + \Delta t) = v_0 \mathcal{N} \left[\frac{1}{N_i} \sum_{j\in S_i} \vs_j + (\hat\eta/v_0) \vxi \right].
\end{equation}
This makes it clear that effectively the interaction term is normalised by the number of interacting particles, which due to the fluctuations in the local environment leads to interactions that can be non-reciprocal.  See \cite{Chepizhko2021} for a detailed discussion of this point.

\subsection{Short Time Dynamics}

\label{sec:short-time-dynamics}

We give here a brief summary of the STD method (reviewed in \cite{Albano_2011}) for the characterisation of phase transitions.  One must first identify the order parameter suitable to describe the transition; in our case this order parameter is the polarisation~\eqref{eq:OP}.  Other quantities monitored to characterise the short time evolution are  the susceptibility $\chi$, the second order Binder cumulant $U_2$, and the logarithmic derivative of the order parameter with respect to  the reduced noise evaluated at the critical point $D$:
\begin{align}
\chi &= \frac {1}{N} (\left\langle \Phi^2\right\rangle-\left\langle \Phi \right\rangle^2), \label{chi} \\
U_2 &=\frac{\left\langle \Phi^2\right\rangle}{\left\langle \Phi \right\rangle ^2} - 1,
\label{U}
\end{align}
where $\tau = (\eta-\eta_c)/\eta_c$ is the reduced noise and $\langle ...\rangle $ indicates average over initial conditions and thermal history. Hereafter, $\left\langle \Phi\right\rangle$ and $\left\langle \Phi^2\right\rangle$ will be referred to as $\Phi$ and 
$\Phi^2$, respectively.

The idea of STD is to analyse the time series of the above observables when the system is initialized with configurations that correspond to trivial fixed points \cite{Albano_2011}.  In the present case, the fixed points correspond to the completely ordered ($\eta=0$) and completely disordered ($\eta=1$) states. In the case of a continuous phase transition, it is expected that at early times of the dynamic evolution, these observables will exhibit a power-law behaviour at the critical point, with exponents that are related to the usual critical exponents of the phase transition. For values of the control parameter $\eta\neq \eta_c$, but within the critical region, the power law is modified by a scaling function. This fact can be used to determine the critical point as well as the critical exponents from the localization of the best power law (for more details see \cite{Albano_2011} and references therein).

For the case of an \emph{ordered} initial condition corresponding to $\eta=0$, the ans\"atze for the time evolution of the observables are
\begin{align}
\Phi(t) &\propto t^{-\beta/\nu z}, \label{STDO} \\
\chi(t) &\propto t^{\gamma/\nu z}. \label{STDchi} \\
U_2(t) &\propto t^{d/z}, \label{STDU} 
\end{align}
where $\beta$, $\nu$, and $\gamma$ are the static critical exponents for the order parameter, correlation length, and susceptibility, respectively; and $z$ the dynamic critical exponent that relates correlation length and correlation time.

If the system is started from a \emph{disordered} initial condition, the scaling laws are instead
\begin{align}
\Phi(t) & \propto t^{\theta},  \label{STDDis} \\
\Phi^2(t) &=\chi(t)\propto t^{\gamma/\nu z}, \label{STDO2}
\end{align}
where $\theta$ is the so-called \emph{initial slip} exponent, encoding the initial increase of the order parameter, and cannot be written in terms of the equilibrium critical exponents.  It is usually expressed in therms of $\beta$, $\gamma$ and $z$ and a new exponent $x_0$.

The universal STD evolution is strictly valid in a-defined time interval $[t_{mic}, t_{max}]$, where the microscopic time $t_{mic}$ is such that the correlation length $\xi(t)$ is of the order of a lattice spacing, and $t_{max}$ is the time when $\xi(t)$ reaches the order of system size $L$.  Furthermore, $t_{max}$ is very small in comparison with the time necessary for equilibration, so that the STD is free from critical slowing down.  In the case of short-range models, the critical point found through STD analysis is free of finite-size effects, since the power laws are determined for times such that $\xi(t)$ remains smaller than $L$ \cite{Albano_2011}.

STD can also be used to find the metastability limits of the phases that coexist in a first-order phase transition \cite{YassJCP2009, Albano_2011}.  In the case of long-range interaction models, the metastability limits are identified with the thermodynamic spinodals.  At these points, the susceptibility and relaxation times diverge, just as in continuous phase transitions, which allows to define them as pseudocritical. In this way, for the ordered initial condition
\begin{equation}
\Phi(t)\propto t^{\omega}+\Phi_{sp}, 
\label{spup}
\end{equation}
here $\omega$ is an exponent and $\Phi_{sp}$ is the value of the order parameter at the spinodal ($\eta^+$). Also, the susceptibility 
diverges as a power law,
\begin{equation}
\chi(t)\propto t^{\Omega}, 
\label{sucepup}
\end{equation}
where $\Omega$ is a different exponent. For the disordered initial condition,
\begin{equation}
\Phi(t)^2\propto t^{\omega^*},
\label{spdown} 
\end{equation}
where $\omega^*$ is another exponent, and $\Phi_{sp}=0$ at the spinodal ($\eta^-$).  At variance with the continuous transition case, the spinodal values do display finite-size effects, as can be seen in Fig.~\ref{fig5} below.

The difference $\eta^+ - \eta^-$ is a measure of the strength of the transition and it allows us to distinguish between a continuous transition (where $\eta^+=\eta^-=\eta_c$, and \eqref{spup} and \eqref{spdown} reduce to  \eqref{STDO} and \eqref{STDO2}, respectively) and a weak first-order transition 
($\eta^- < \eta_c <\eta^+$, where $\eta_c$ is the value of the control parameter at coexistence \cite{Albano_2011}).

\subsection{The coefficient of determination}
\label{rmethod}

To apply  STD to find the critical point then one has to find the control parameters so that the order parameter, cumulant, etc.\ exhibit the longest-running power law: these are the critical parameters.  Here we use a quantitative way of finding the best power law, employing a refinement technique proposed in \cite{silva2020}.  One defines the coefficient of determination $r$ \cite{silva2020}  by
\begin{equation}
r=\frac{\sum\limits_{t=N_{\min }}^{N_{MC}}(\overline{\ln \Phi}-a-b\ln t)^{2}}{%
\sum\limits_{t=N_{\min }}^{N_{MC}}(\overline{\ln \Phi}-\ln \Phi(t))^{2}},
\label{eq:coef_det}
\end{equation}
where $N_{MC}$ is the total number of MC steps and $\overline{\cdots}$ stands for the time average over the number of steps included in the fit,
\begin{equation}
\overline{\ln \Phi}=\frac{1}{(N_{MC}-N_{\min }+1)}\sum\nolimits_{t=N_{\min
}}^{N_{MC}}\ln \Phi(t)\text{.}
\end{equation}
The value of $N_{\min}$ depends on the details of the system under study and is related to the microscopic time scale, i.e., the time the system needs to
reach the universal behaviour in short-time critical dynamics \cite{janssen1989}.  The denominator of \eqref{eq:coef_det} represents the dispersion, or variation, of the data around the time average, while the numerator is the \emph{explained} variation, i.e.\ the dispersion accounted for by the fit.  The value of the coefficient of determination is $r\le1$, with $r=1$ holding for a fit that can explain all the variation (a perfect fit) (see appendix of ref.~\cite{silva2020} for details).

When the system is at criticality, for an initial condition $\Phi_{0}=1$ ($\Phi_{0}=0$), we expect that the order parameter follows a power law $\overline{ \Phi(t)}\sim t^{-\beta /\nu z}\ $ ($\overline{
\Phi(t)}\sim t^{-\theta}\ $), which gives $r=1$ for $N_{MC}\to\infty$.  On the other hand, when the system is out of criticality, there is no power law and $r < 1 $.  Thus the critical value of $\eta$ is determined as the value that gives the largest value of $r$.

The power-law fit is done at fixed $N_{MC}$ and, for the case of \eqref{spup}, $\Phi_{sp}$.  To find these values we also employ the coefficient of determination, choosing the values of $N_{MC}$ (and $\Phi_{sp}$, if applicable) that give the highest $r$ at a given fixed $\eta$.

Due to its non-universal nature, $N_{\min}$ is fixed somewhat arbitrarily, but its value is the same for all values of the control parameter.  In practice, one discards the first $N_{\min }$ MC steps which obviously deviate from a power law.  We have checked that small variations of $N_{\min}$ do not alter the determination of the critical point.

\section{Results and Discussion}

\subsection{Short-time dynamics of the static Vicksek model}
\label{sec:aclstd-static-aclvm}

We first apply STD to the static VM, where the aligning (``ferromagnetic'') interaction \eqref{vm1} is applied but the position update \eqref{vm2} is disregarded, and the vector $v$ reduces to an internal (``spin'') degree of freedom.  Since the particles are not free to move, to completely define the model, the particles' positions must be specified, which in turn fixes the interaction graph through the cut-off distance, $r_c=1$.  We have considered two different geometries:
\begin{itemize}
\item a simple cubic (SC) lattice, and
\item a Poisson Random Graph.
\end{itemize}
In the SC lattice each particle is located at  $(i,j,k)$ with $i,j,k \in \mathbb{N}$.  Each particle then interacts with its 6 nearest neighbours,  and the linear size of the lattice is $L=N^{1/3}$.  The Poisson graph is generated by placing $N$ particles randomly, independently and uniformly distributed in a cubic box of side $L$, and then connecting the particles within the cut-off radius.  The number of particles within the interaction sphere of any given particle is Poisson-distributed with mean $\approx 4.18$.  It can happen that the random graphs generated in this way have disconnected components, i.e.\ that there exist pairs of nodes such that there is no path along the graph's edges that connects them.  This is a hindrance regarding the achievement of global order, since the disconnected components will order separately and almost surely point in different directions  when breaking the symmetry.  However, this is not a severe concern at the relatively high density we use here; we have found that the largest connected component of the graph contains more than 95\% of particles in all the samples used in the simulations.

The static VM is sometimes referred to as the $v_0\to0$ limit, although to implement the simulation one keeps $v_0\neq0$ to define a local direction (the actual value of $v_0$ is irrelevant).  More importantly, just stating that the speed tends to 0 does not completely specify the model; one should specify the path in the $(\eta, \rho, v_0)$ parameter space one chooses to reach $v_0=0$.  This is because any order in the velocities will be reflected in some (inhomogeneous) positional structure.  Our Poisson graph setup could be obtained by letting $v_0\to0$ at fixed density and very high value of the noise, so that the system remains completely disordered.  On the other hand, the SC lattice is an arbitrary choice that is not generated by letting $v_0$ tend to zero in an actual simulation of the model.

We study the order-disorder transition using the STD procedure as outlined in Sec.~\ref{sec:short-time-dynamics}.  Our aim is to establish the validity of the STD analysis in this mildly out-of-equilibrium situation where detailed balance is not strictly obeyed.  We consider both PU and SU dynamics. We have used $\rho=N/L^3 = 1$ and $N=4913$, 10648, and 19836, corresponding to $L=17$, 22, and 27.

\begin{figure}[!h]
\centering
\includegraphics[width=0.8\columnwidth]{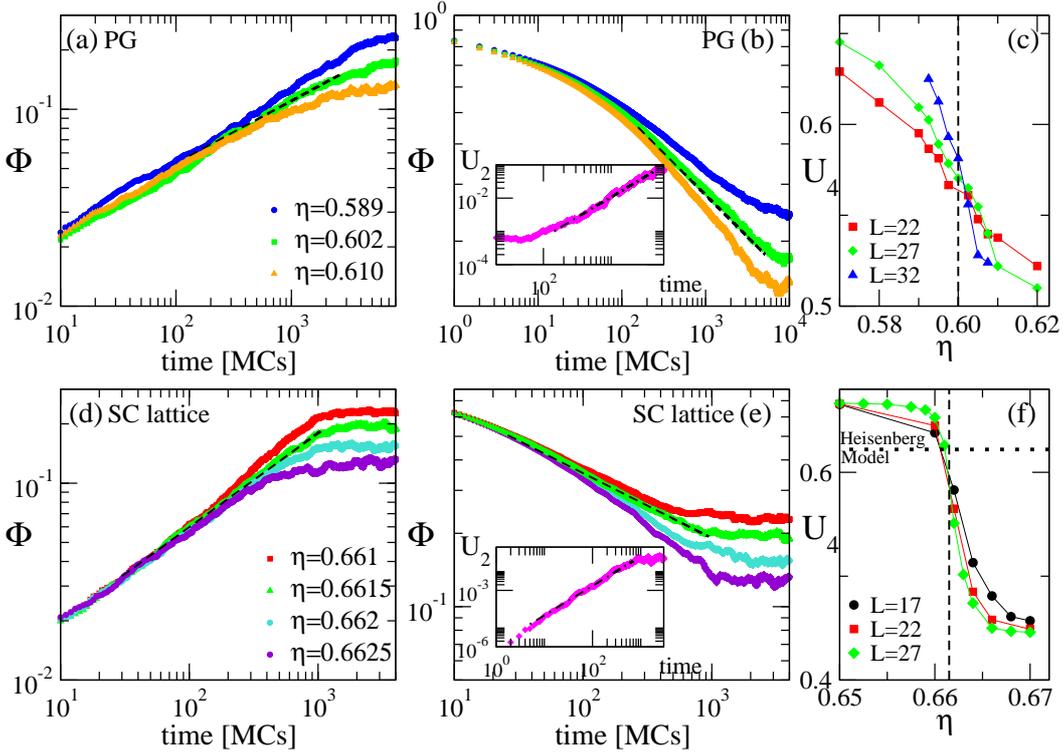}
\caption{Typical STD evolution of the order parameter for the static VM with PU on the SC lattice and on the Poisson random graph.  Curves are averaged over 100 independent realisations of the dynamics and, where appropriate, disorder, for $N=19863$, $\rho=1$ ($L=27$), and different values of noise $\eta$ as indicated. \emph{(a)} and \emph{(b):} Poisson graph with DICs and OICs, respectively. Black dashed lines are the power law fit, yielding the critical noise $\eta_c=\eta^+=\eta^-=0.604(1)$. The slopes are $\theta=0.32(1)$ for DICs, and $\beta/\nu z = 0.25(1)$ for OIC. Inset of figure \emph{(b)}: Evolution of the second order Binder cumulant $U_2$ \eqref{STDU} at the critical noise $\eta_c=0.604(1)$ for the Poisson random graph. Dashed line has slope $d/z=1.22(1)$. Note that for the largest times the system is approaching equilibrium and the value of $\Phi$ reached for both initial conditions is similar.  For those time the system is outside the short-time regime.
\emph{(c)} Fourth-order Binder cumulant ($U_4$) vs $\eta$ for different system sizes $L$, as indicated. Black dashed line indicates $\eta_c^{STD}\simeq 0.60$.
\emph{(d)} and \emph{(e)}: SC lattice with DICs and OICs respectively.  Black dashed lines are power law fits giving $\eta_c=\eta^+=\eta^-=0.6615(5)$. For the exponents we obtain $\theta=0.48(1)$, and $\beta/\nu z = 0.27(1)$. Inset of figure \emph{(e)}: Time evolution of $U_2$ \eqref{STDU} at the critical noise $\eta_c=0.6615(5)$ for the SC lattice. Dashed line has slope $d/z=1.50(1)$.  \emph{(f)} Fourth-order Binder cumulant ($U_4$) vs $\eta$ for different system sizes $L$, as indicated. Dashed line indicates $\eta_c^{STD}\simeq 0.661$, and dotted line corresponds to the fixed-point cumulant value ($U^*=0.622(1)$) determined in the Heisenberg model \cite{peczak}.}
\label{fig1}
\end{figure}

We have recorded the evolution of the order parameter and the second-order Binder cumulant $U_2(t)$ starting from ordered and disordered initial conditions, i.e.\ all velocities pointing in the same direction, and each velocity independently assigned a random orientation, respectively.  The short time behaviour evolution in this case is qualitatively the same as found in equilibrium systems.  Typical curves for the evolution of the order parameter are shown in Figure~\ref{fig1}.  A reasonably wide power law in time can be found for a range of values of the noise $\eta$.  From these curves, $\eta^+$ and $\eta^-$ are obtained as the values of $\eta$ which yield the longest power law when starting from ordered initial conditions (OICs) or disordered initial conditions (DICs), respectively.  For this case both values were coincident, and no significant size effects were observed.  This holds for both kinds of update and for both graphs.  This observation indicates a continuous transition, as in the Heisenberg model.  STD then yields the critical value $\eta_c=\eta^+=\eta^-$, as well as the critical exponents, as summarised in Sec.~\ref{sec:short-time-dynamics}.

Qualitatively, this finding is expected, but we check it quantitatively by determining the critical noise through the finite size behaviour of the fourth-order cumulant $U_4 \equiv 1-\frac{\left\langle \Phi^4\right\rangle}{3\left\langle \Phi^2 \right\rangle ^2}$.  We did some of simulations of systems with $L=17$, 22, 27, and 32 running long enough ($t\simeq 10^6$) so that the system reaches a stationary state.  It is well known that, in the case of a continuous phase transition, the fourth-order cumulant $U_4(L) \rightarrow 4/9$ in the disordered phase,  $U_4(L) \rightarrow 2/3$ in the ordered phase, and at the critical point $U_4(L) \rightarrow U^*$ in the limit $L\rightarrow \infty$.  As the rightmost panels of Figure~\ref{fig1} show, the cumulant confirms the continuous phase transition scenario, and the critical noise found from the crossing of the cumulants for different sizes coincides with the $\eta_c$ obtained from STD.  For the SC lattice, the value of the $U^*$ of the cumulant at $\eta_c$ is very close to the value  $U^*=0.622(1)$ reported for the Heisenberg model on the same lattice \cite{peczak}, where the critical coupling is $K_c=J/k_B T_C\simeq 0.693$.

The critical exponents obtained from the STD of $\Phi(t)$ and $U_2(t)$ for PU dynamics in both graphs are shown in Table~\ref{tab:exponents-PU}, together with the corresponding values for the equilibrium Heisenberg model.  The exponents found for the SC lattice are compatible with those of the Heisenberg model on the same graph, but on the Poisson graph the exponents take different values.  This is not surprising, since not only the number of nearest neighbours is fluctuating on this graph, but also its mean is significantly smaller than in the SC case.  

\begin{table}
  \caption{Critical noise and exponents for the static Vicsek model in the SC and Poisson graphs with PU as obtained from STD.  For comparison, the values for the 3-$d$ Heisenberg model from refs.~\cite{Fernandes_2006,Albano_2011} are quoted.} 
  \begin{center}
  \begin{tabular}{@{}llllll}
\hline              
    & SC lattice  & Poisson graph & 3-$d$ Heisenberg\\
\hline 
    $\theta$      & 0.48(1) & 0.32(1)  & 0.482(3) \\
    $\beta/\nu z$ & 0.27(1) & 0.25(1)  & 0.266(3) \\
    $z$           & 2.00(1) & 2.46(2)  & 1.976(9) \\
    $\eta _c$     & 0.662(1) & 0.604(1)
  \end{tabular}
\end{center}
\label{tab:exponents-PU}
\end{table}

From these results we conclude that STD can be applied to the static VM despite the lack of detailed balance, and that the Heisenberg model can be considered the static limit of the VM.

We have also measured the exponents in the case of SU dynamics (Table~\ref{tab:exponents-SU}).  In a Monte Carlo simulation of an equilibrium model, one would not expect that such a change make a difference.  However, in the static VM, the situation depends on the geometry: on the SC lattice, the exponents for SU or PU dynamics are the same.  On the other hand for the Poisson graph there is a noticeable difference between the two dynamic rules.  This sensitivity to the dynamics in the Poisson graph case may be related to the fact that the interaction is non-reciprocal in this case: as mentioned in Sec.~\ref{sec:detailed-balance}, the factor $N_i$ in the noise rule leads to non-reciprocal interactions when the number of neighbours can fluctuate locally, which happens in the Poisson graph, but not in the SC lattice.

\begin{table}
  \caption{Critical noise and exponents for the static Vicsek model in the SC and Poisson graphs with SU from STD.}
\begin{center}
\begin{tabular}{@{}llllll}
\hline                          
 & SC & Poisson graph \\
\hline 
$\theta$      & 0.47(1)  & 0.37(1)  \\
$\beta/\nu z$ & 0.27(1)  & 0.18(2)  \\
$z$           & 2.00(1)   & 2.46(1)  \\
$\eta _c$      & 0.656(1)  & 0.465(1)   \\
\end{tabular}
\end{center}
\label{tab:exponents-SU}
\end{table}

\subsection{Short-time dynamics of the active Vicsek model}
\label{sec:aclstd-active-aclvm}

Now we turn our attention to the STD of the active VM.  We shall consider a system with density $\rho$=1 and speed $v_0=0.5$.  In this case the transition is first order \cite{ginelli_physics_2015}, but its discontinuous character is weaker the smaller the system size, and it can appear as continuous for small systems and low speed \cite{chate2008} (the effect is even more marked for angular noise than for the vector noise we use here).  In an equilibrium first-order transition, the STD method of finding the best power law does not find the transition.  Rather, it yields two different values of the control parameter, depending on whether the initial condition is ordered or disordered.  These values bracket the critical control parameter, and are closer to each other the weaker the transition \cite{schulke_dynamic_2000}.  It has been shown \cite{YassJCP2009} that they can be interpreted as the spinodal points of the first-order transition.  Here we aim to check whether this scenario still holds in the active case.

To prepare the systems with disordered initial conditions (DICs), positions and velocity where assigned independently and randomly with a distribution uniform in the simulation box and in all solid angles, respectively.  The STD when starting from disorder follows the usual equilibrium STD phenomenology, with the control parameter growing as a (transient) power law, which lasts for a longer time at a particular value of $\eta$.  This holds both for the standard PU and for SU.  In the SU case the duration of the power law is shorter, and the order parameter seems to approach equilibrium faster than when using PU.  The results that follow concern the standard PU dynamics, but this observation turns out to be useful for the preparation of the system under ordered initial conditions (OICs).

OICs are not easily prepared by hand due to the strong density-velocity coupling characteristic of the VM and at the heart of several of its particular properties, like the fact that the model can achieve long-range order in 2-$d$ \cite{toner_flocks_1998, ramasw}: simply aligning the velocities of a set of randomly-placed particles would generate a configuration that is ordered but atypical.  For this reason we prepare a proper ordered state by starting with a random configuration and letting the system to evolve with zero noise until it reaches a fully polarised state ($\Phi=1$).
 
\begin{figure}[!h]
\centering
  
    \includegraphics[width=0.8\columnwidth]{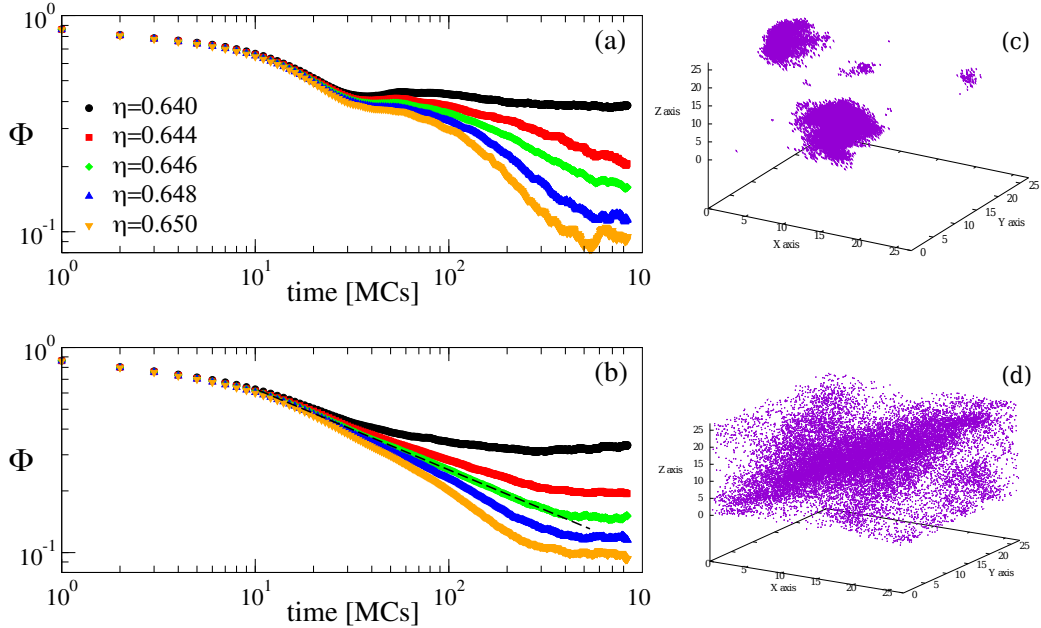}
 
  \caption{STD of the active VM.  A system of size $L=5000^{1/3}\approx 17.1$ was prepared with OIC as described in the text.  The curves shown in both panels correspond both to the standard VM with vector noise and PU.  Different dynamics were used during the preparation of the initial configuration: \emph{(a)} PU or \emph{(b)}  SU.  Curves are averaged over 2000 samples.  In the system prepared with PU, an anomalous decay is found, with a local minimum and without a clear initial power law. 
  Snapshots configurations obtained after $t=10^4$mcs, for $\eta=0$, $N=27000$, and with \emph{(c)} PU and \emph{(d)} SU, characteristics of the initial preparation used to analyze the STD starting from an ordered initial condition (OIC).}
\label{fig2}
\end{figure}

Typical STD curves of systems prepared in this way are shown in Figure~\ref{fig2}.  Here the short-time behaviour of the OP is anomalous: instead of a monotonic decay, a local minimum appears near $t\approx 50$, and no clear power law is discernible (Figure~\ref{fig2}\emph{(a)}).  

If instead we prepare the OIC using \emph{sequential} update (but then run with PU as is standard), we recover the qualitative STD behaviour one expects in equilibrium: the decay is now monotonic, and a power-law regime can be found (Figure~\ref{fig2}\emph{(b)}). 
We interpret this behaviour as the SU being more efficient in finding a configuration of particle positions representative of the low-temperature phase (more on this below). 
Right panels of Figure~\ref{fig2} show typical snapshot configurations obtained when the system, started from a disordered initial configuration, evolves after $t=10^4$mcs, and with zero noise, with PU (Figure~\ref{fig2}\emph{(c)}), and SU (Figure~\ref{fig2}\emph{(d)}), respectively.
Even when, in both cases, the final state is fully polarised ($\Phi=1$), one can see that SU dynamics produces band-like configurations, whereas with PU one obtains essentially a blob of ordered particles.  In this sense, the configurations obtained with SU seem to be more appropriate as the initial ordered state to the STD analysis in the case of OIC.

\begin{figure}[!h]
\centering
   \includegraphics[width=0.7\columnwidth]{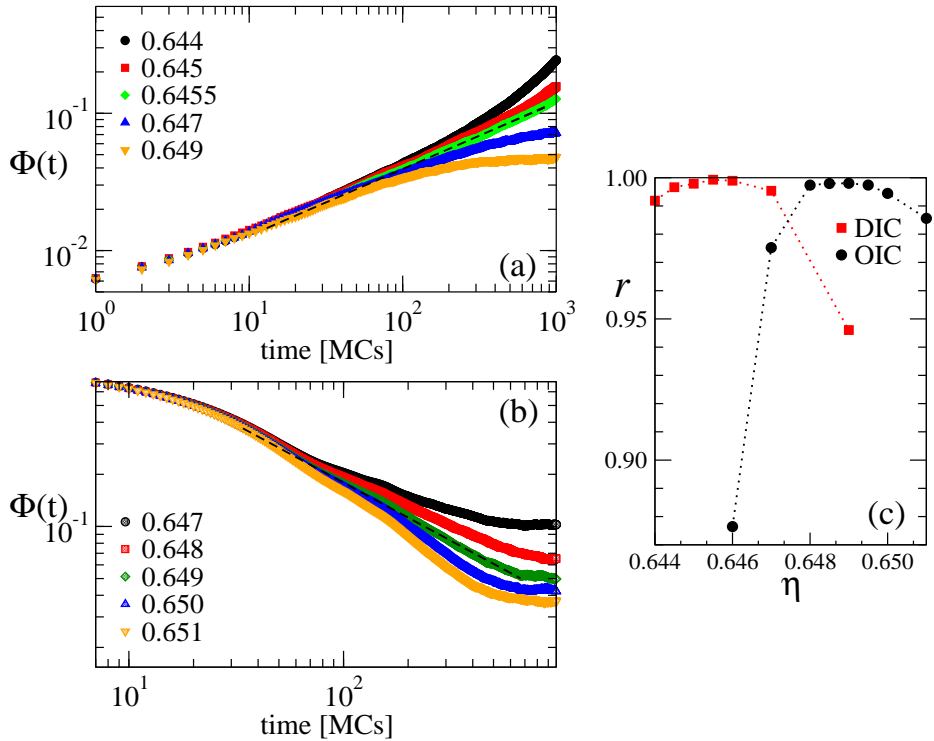}
   
   \caption{STD of the active VM for size $N=27000$ ($L= 30$ and with $n_{sample}=1000$, by starting the system with \emph{(a)} disordered initial conditions (DICs), and \emph{(b)} ordered initial conditions (OICs). Panel \emph{(c)} shows the results of the $r$ analysis (\ref{eq:coef_det}). The maximum values of $r$ indicate the $\eta$ of the best power law for the curves in \emph{(a)} and \emph{(b)}.}
  \label{fig3}
\end{figure}

With this protocol we can then successfully find a power-law regime (Figure~\ref{fig3}\emph{(a)-(b)}) and find the values $\eta^+$ and $\eta^-$ that correspond to the best power laws starting from DICs and OICs respectively (Figure~\ref{fig3}\emph{(c)}).  The $\eta^+$ and $\eta^-$ found in this way for different system sizes are reported in Figure~\ref{fig5}.

\begin{figure}
  \centering
  \includegraphics[width=0.7\columnwidth]{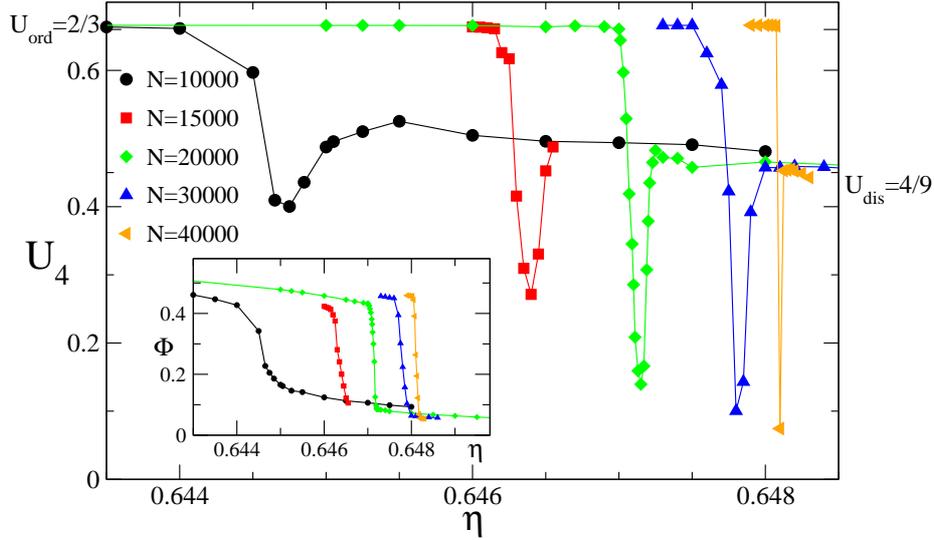}
  \caption{Fourth-order cumulant $U_4$ as a function of the noise $\eta$, for different number of particles $N$. The number density is $\rho=1.0$ and the particle velocity is $v_0=0.5$. The values expected to $U_4 \rightarrow 4/9=U_{dis}$ in the disordered phase,and $U_4\rightarrow 2/3 =U_{ord}$ in the ordered phase are indicated. Inset: Order parameter vs $\eta$.}
  \label{fig4}
\end{figure}

The fact that sequential update is needed for the preparation of the OIC can be rationalised taking into account phase coexistence, which happens at values of $\eta$ intermediate between those corresponding to full order or full disorder.  More precisely, the phase diagram in the $(\rho, \eta)$ plane (see e.g.\ Fig.~1 of reference \cite{Chate2020}) has two homogeneous phases (disordered gas and orientationally ordered liquid), separated by a coexistence region where ordered bands are divided by disordered gas-like zones.  Since we are using polarisation as order parameter, STD is sensitive to the onset of order, which happens when bands start to form, i.e.\ at the transition between gas and liquid-gas coexistence.  Thus the spinodals $\eta^+$ and $\eta^-$ found here bracket the gas-to-coexistence transition.  The OIC then, should be prepared in the coexistence, rather than in the homogeneous fully ordered phase.  A similar situation was found in a study of the liquid/gas spinodals in a Lennard-Jones liquid \cite{loscar_spinodals_2016}, where starting with OIC actually meant starting from the low-temperature side of the transition, i.e.\ the liquid.  The difference between using PU or SU to prepare the OIC is precisely that with SU one obtains a configuration with a band-like structure, while PU leads to a completely ordered and inhomogeneous configuration, as witnessed by the snapshots in Fig.~\ref{fig2}.  The picture emerging from the STD study is then that of a system with a discontinuous ordering transition.  This is fully compatible with what is known about the VM, indicating that the STD can be successfully applied to this active system.  Our $\eta^-$ corresponds to the gas spinodal (called $S_\text{gas}$ in ref.~\cite{Chate2020}), i.e.\ the point where the disordered phase becomes unstable.  As for $\eta^+$, it is the point where coexistence becomes unstable.  In the thermodynamic limit, this should coincide with $\eta_c$, the critical value for the onset of order determined in the stationary state.  Indeed, Fig.~\ref{fig5} below is compatible with $\eta^+$ merging with $\eta_c$, the latter having been determined as explained next.

As in the static case, we have checked quantitatively the results of STD by computing the critical value of the noise at the ordering transition, $\eta_c$, using standard finite-size scaling of the Binder cumulant, as in ref.~\cite{chate2008}.  We therefore simulated several $5000 < N < 40000$ and with $v_0=0.5$ and $\rho=1$ as before.  Runs lasting up to $t=10^6$ to $5\cdot10^6$ were used, discarding a first transient stage of $2\cdot 10^5$ to $10^6$ time units.  In this way, the average order parameter $\Phi$ in the steady state was measured as a function of noise.  We used two sets of several (typically four) independent runs, starting from OICs and DICs.  The reported values of the order parameter are averages over all runs.  We checked that within statistical errors the values obtained with each initial condition coincide.  We also computed the second and fourth moments of the order parameter, and the fourth-order cumulant $U_4$.

The $\eta$-dependence of the order parameter (Figure~\ref{fig4}) shows how the transition grows sharper, and more clearly discontinuous, as $L$ (or equivalently $N$ since the density is fixed) is increased.  Starting from the ordered phase at low $\eta$, for sizes up to $N \approx 10000$ a smooth decay is obtained, similar to a continuous transition.  However, when $N$ is increased further, a steeper transition, clearly discontinuous, is observed.  The jump of $\left\langle \Phi \right\rangle$ across the transition increases slightly with size, a typical size effect observed in simulations of thermodynamic systems in equilibrium.
	
A more stringent probe of the nature (continuous/discontinuous) of the transition is obtained with the fourth-order Binder cumulant (Figure~\ref{fig4}).  This quantity develops a clear minimum at the transition, which becomes sharper and deeper as size is increased.  This is a clear signature of a discontinuous transition.  The critical value of the noise $\eta_c$ extracted from the minimum of the cumulant is plotted as function of size in Figure~\ref{fig5}.  $\eta_c$ lies in between the values $\eta^+$ and $\eta^-$ obtained from STD with OICs and DICs, as observed in equilibrium first-order transitions \cite{schulke_dynamic_2000, YassJCP2009}.  The size dependence of $\eta_c$ as well as $\eta^+$ and $\eta^-$ is found to be compatible with an $L^{-d}$ correction, i.e.\ $\eta_c(L) = \eta_c(\infty) + a L^{-d}$ (Fig.~\ref{fig5}, inset), as in equilibrium first-order transitions.  Thus the finite-size scaling of the cumulant (in equilibrium) confirms the STD results both qualitatively and quantitatively.
Additionally, $\eta^+$ tends to coincide with $\eta_c$ when $L\to\infty$.

\begin{figure}[!h]
\centering
    \includegraphics[width=0.6\columnwidth]{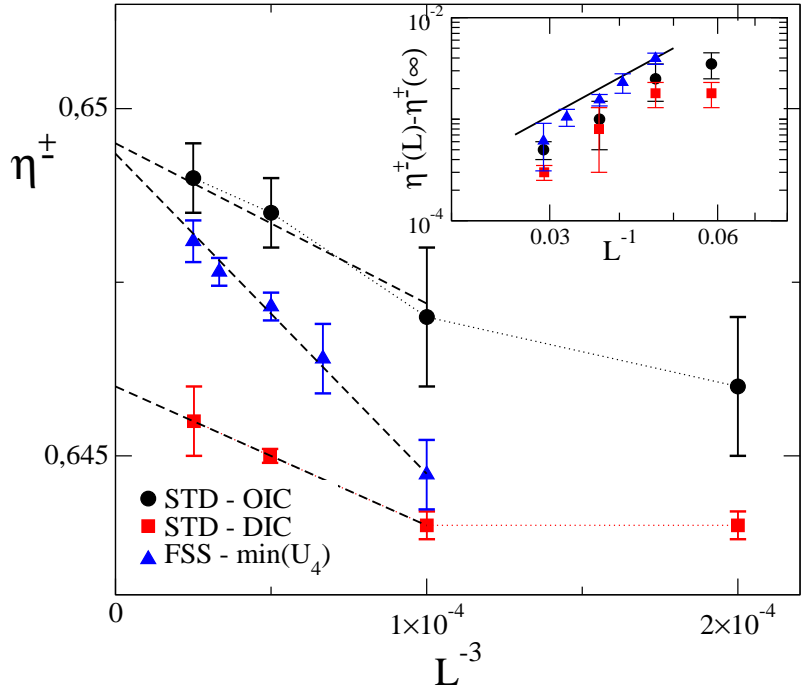}
  
  \caption{$\eta^+$ and $\eta^-$ as a function of $L^{-d}$ ($d=3$). The critical value of the noise $\eta_c$ extracted from the minimum of the cumulant $U_4$ are in between the values $\eta^+$ and $\eta^-$ obtained from STD with OIC and DIC, as expected.  For $L\to\infty$ both $\eta_c$ and $\eta^+$ approach the same value. 
  Inset: Size dependence of $\eta_c^{+_-}(L) - \eta_c^{+_-}(\infty)$ as s function of  $L^{-1}$. Full-line has slope $=3$, which confirms the $L^{-d}$ behaviour observed.}
  \label{fig5}
\end{figure}

\section{Conclusions}
\label{sec:conclusions}

We have studied the short-time dynamics behaviour of the Vicsek model, one of the simplest and perhaps the most well-known active matter model.  The aim was to check whether the STD in active models is similar to the behaviour of equilibrium systems near criticality \cite{janssen1989} or spinodal points \cite{YassJCP2009}.

Our results indicate that the STD in this active model is qualitatively very similar to the equilibrium STD.  In the static ($v_0\to0$) case, where the transition is continuous (although detailed balance is not strictly obeyed), the STD displays the longest power-law in time at the critical noise, and allows to determine the critical exponents.  Determination of critical exponents requires following the STD starting from both order and disorder, but preparation of these initial conditions is straightforward in the static case.

In the active case the transition is no longer continuous, and STD behaves as in first-order equilibrium transitions, yielding two values of the noise, $\eta^+$ and $\eta^-$, depending on whether the initial condition is ordered or disordered, that bracket the transition and are further apart for larger systems, where the transition is more strongly discontinuous.  In this case we have found that preparation of the ordered initial conditions is tricky: to obtain a power law, we have prepared the OIC by running an initially disordered configuration at $\eta=0$ but with \emph{a different dynamics} (sequential update) than the paralell update that is used by definition in the standard VM.  This can be understood taking into account that homogeneous order and disordered phases are separated by a coexistence region: the OIC conditions for STD actually mean starting from the low temperature phase just next to the disordered phase, i.e.\ the phase-separated one.  This is what is achieved by using the PU dynamics.  The value $\eta^-$ marks the gas spinodal, where the disordered phase loses stability, this is called $S_\text{gas}$ in ref.~\cite{Chate2020}.  The liquid spinodal, corresponding to the homogeneous ordered liquid losing stability in favour of phase separation, is not seen here, since for our initial conditions, near to the onset of order, the STD brackets the $\eta_c$ where phase separation starts to be seen.

The present results show that STD can be used as a tool, as in equilibrium matter \cite{Albano_2011} to study the characteristics of the VM order-disorder transition, and suggest that STD can be a very useful technique to investigate active systems without need to reach a stationary state.  Also, from the critical exponents determined for the static case, one can conclude that the static VM can be regarded as a kind of Heisenberg model despite the lack of strict detailed balance.

Although investigation of other active models is needed in order to confirm that the present results conform to a general picture in active matter, these results open the way to the use of STD to study the phase diagram of the Vicsek model, and suggest that it is a very useful technique to add to the toolbox of researchers exploring active systems.  An obvious but interesting direction to pursue is to attempt to use STD to study the loss of stability of the homogeneous ordered phase (the $S_\text{liq}$ spinodal of ref.~\cite{Chate2020}).  This will require to identify an appropriate order parameter.

\section*{Acknowledgements}

We thank an anonymous reviewer for useful comments on the original manuscript. This work was supported Consejo Nacional de Investigaciones Cient\'ificas y T\'ecnicas (CONICET), Universidad Nacional de La Plata (Argentina), and Agencia Nacional de Promoci\'on de la Investigaci\'on, el Desarrollo Tecnol\'ogico y la Innovaci\'on (Agencia I+D+i). Simulations were done on the cluster of Unidad de C\'alculo, IFLYSIB (Argentina).
\\

\bibliographystyle{unsrt}
\bibliography{bibRP}

\end{document}